\documentclass{svproc}
\usepackage{graphicx}
\usepackage{nameref}
\usepackage{blkarray,bigstrut}
\graphicspath{ {./images/} }
\usepackage{cite}
\usepackage{float}
\usepackage{physics}
\usepackage{amsmath}
\usepackage{amsfonts}
\usepackage{comment}
\usepackage{lscape}
\usepackage{indentfirst}
\usepackage{latexsym}
\usepackage{multirow}
\usepackage{tabls}
\usepackage{wrapfig}
\usepackage{longtable}
\usepackage{supertabular}
\usepackage{subfigure}
\usepackage{algpseudocode}
\usepackage{fancyvrb}
\usepackage{afterpage}
\usepackage{subfig}
\usepackage{url}

\begin{document}
\mainmatter              
\title{Quantum Optimized Centroid Initialization}
\titlerunning{Quantum Optimized Centroid Initialization (QOCI)}  
%
\author{Nicholas R. Allgood\inst{1} Ajinkya Borle
Charles K. Nicholas}
\authorrunning{Nicholas R. Allgood et al.} 
%
\tocauthor{Nicholas R. Allgood, Ajinkya Borle, Charles K. Nicholas }
\institute{University of Maryland Baltimore County, Baltimore, MD 21250 USA
\email{allgood1@umbc.edu, aborle1@umbc.edu, nicholas@umbc.edu}}

\maketitle              

\begin{abstract}
One of the major benefits of quantum computing is the potential to resolve complex computational problems faster than can be done by classical methods. There are many prototype-based clustering methods in use today, and the selection of the starting nodes for the center points is often done randomly. Clustering often suffers from accepting a local minima as a valid solution when there are possibly better solutions. We will present the results of a study to leverage the benefits of quantum computing for finding better starting centroids for prototype-based clustering.
\end{abstract}

\section{Introduction}
Clustering of data must start from somewhere, and the well-known $k$-means and $k$-medoids are no exception. These methods of clustering select a center point to start grouping related data around them to form their clusters. In much of the literature on the subject, we find that the center points to start with are selected at random. To get the best and most accurate groups of related data, we must have accurate center points to formulate a cluster, whether it be a computed mean or a sample from the data. Many clustering methods employ an iterative verification of the selected center point to confirm it is the best candidate to be the centroid. \cite{algorithms_clustering_data}\\

We could perhaps save some computational overhead if we begin the clustering algorithm with the \textit{best} possible center nodes computed prior to forming the cluster. In classical cases, this could be quite computationally expensive, significantly more so than using a random process to select the starting nodes. With the advantages of quantum computing, we could theoretically offload this process to a quantum computer, which would then calculate the best possible center nodes. A major benefit to doing this, other than potential performance increases, is even if a performance increase isn't found, we immediately know that we have the best possible set of center nodes and we simply have to continue forming the clusters by grouping pieces of related data. A final point is that we can take some comfort in knowing that with a quantum process, we are getting a true global minimum based on the entire sample data and avoid the well-known \textit{local minima}.

\section{Related Work}

Cluster analysis is a class of algorithms that group a set of objects in such a way that objects in the same group are more similar to each other than to those in other groups. These groups of data are known as \textit{clusters} and there are a wide variety of metrics that can be used to determine how similar an object is to another object. Cluster analysis is widely used in statistical data analysis for pattern recognition, image analysis, and machine learning and countless other domains.

\subsection{K-Means}

$k$-means clustering is a method of grouping related data together, more formally known as vector quantization\cite{MacQueen_clustering}. The idea is to partition $n$ observations into $k$ clusters where each observation belongs to the cluster with the closest mean (also known as the cluster center or \textit{centroid}). This serves as the base or prototype for the cluster, which in turn implies that $k$-means and others belong to a family known as \textit{Prototype-Based Clusters}. The data is partitioned such that variance inside of a cluster are minimized. It should be noted that a key difference with $k$-means utilizes squared Euclidean distances instead of regular Euclidean distances. The latter where regular Euclidean distances are used is a much harder problem known as the Weber problem.\cite{weber_prob}  

While there are many variants and formulations of $k$-means, we will be discussing only  the standard version known as naive $k$-means. Mathematically, $k$-means is defined as follows: given a set of observations ($x_1$, $x_2$, ..., $x_n$), we partition the $n$ observations into $k \leq n$ sets $S = \{S_1, S_2, ..., S_k\}$ so to minimize the variance within each of the sets $S_i$. Formally:

\begin{equation}
    \min_s\sum_{i=1}^{k}\sum_{x \in S_i} ||x - \mu_i||^2 = \min_s \sum_{i=1}^{k} |S_i|   \mathrm{Var} (S_i) 
\end{equation}

where $\mu_i$ is the mean of the points in $S_i$. 

The full algorithm consists of three pieces, namely initialization, assignment, and update.  
\begin{enumerate}

\item{\textbf{Initialization}}

Let us randomly choose $k$ observations from our data sets and use these as our initial centroids $m_1^{(1)}, ..., m_k^{(1)}$


\item{\textbf{Assignment}}
Given our initial centroids $m_1^{(1)}, ..., m_k^{(1)}$, assign each observation to the cluster with the nearest centroid in terms of least squared Euclidean distance.

\begin{equation}
    S_i^{(t)} = \big \{ x_p : ||x_p - m_i^{(t)}||^2 \leq ||x_p - m_j^{(t)}||^2\ \forall j, 1 \leq j, \leq k \big \},
\end{equation}
where every $x_p$ is assigned to exactly one $S^{(t)}$, even when the possibility of being assigned to multiple $S^{(t)}$ exists.

\item{\textbf{Update}}

This step recalculates the centroids (means) for observations to each cluster:

\begin{equation}
    m_i^{(t+1)} = \frac{1}{|S_i^{(t)}|} \sum_{x_j \in S_i^{(t)} x_j}
\end{equation}

The algorithm has fully converged when the assignments no longer change. While there are variants that address this, one major shortcoming of naive $k$-means is there is no guarantee to find a global optimum. 

\end{enumerate}

\subsection{Quantum Annealing}

Quantum annealing is a method of quantum computing that relies on the adiabatic principal of the evolution of a quantum state over time. This is formally expressed as:
\begin{equation}
    i\hbar\frac{d\ket{\psi}}{dt} = H\ket{\psi}
\end{equation}
where $t$ is time, $\psi$ is the state of the system, and $\hbar$ is Planck's constant for which the exact value is not known, but for many practical purposes $\hbar = 1$ suffices.

A Hamiltonian is a mathematical model for expressing the total energy of a system and quantum annealing problems are expressed as a Hamiltonian with a minimization of energy as $H_p$. Once processed over time, our solution is an eigenvector with the lowest energy in $H_p$. The adiabatic principle states that as our time $T$ tends towards $\inf$, we will be very close to the ground state of energy in $H_p$. 

Quadratic unconstrained binary optimization (QUBO) is a combinatorial optimization problem that is NP-hard and has a wide variety of applications such as machine-learning, economics, and finance. The set of binary vectors of a fixed length $n>0$ is denoted by ${B}^{n}$, where $B = \{0, 1\}$ is a set of binary values. Given a real upper triangular matrix $Q \in \mathbb{R}^{nxn}$ whose entries $Q_{ij}$ define a weight for a pair of indices $i,j \in \{1, \dots, n\}$ within a binary vector. A function $f_Q: \mathbb{B}^n \mapsto \mathbb{R}$ assigns a value to each binary vector:

\begin{equation}
    f_Q(x) = x^TQx = \sum^{n}_{i=1}\sum_{j=1}^{n}Q_{ij}x_{i}x_{j}
\end{equation}

With a QUBO problem, we wish to find a binary vector $x^*$ that is minimal to $f_Q$ such that:
\begin{equation}
x^* = \arg\min_{x\in\mathbb{B}^n}f_Q(x)
\end{equation}

Quantum annealers such as those used by D-Wave \cite{dwave}, rely on a formulation of problem as an Ising Hamiltonian or a Quadratic Binary Optimization Problem (QUBO) both of which are equivalent expressions of a problem.

\subsection{Non-Negative Matrix Factorization}

Non-negative matrix factorization (NMF) are a group of algorithms where a matrix $V$ is factorized to two matrices $W$ and $H$ where all three matrices have no negative elements. While generally the problem is not usually solvable, it is often numerically approximated. NMF also has a wide variety of fields such as astronomy, computer vision, document clustering, signal processing, and many others. 

Matrix multiplication is implemented by computing the columns of vectors $V$ as a linear combination of the column vectors in $W$ with the coefficients supplied by the columns in matrix $H$. We are then able to state that each column of V can be computed as $v_i = Wh_i$ where $v_i$ is the $i$-th column vector of the product matrix V and $h_i$ is the $i$-th column vector of the matrix $H$. The multiplication of the matrices will often result in dimensions of $W$ and $H$ being much lower than those of the product matrix. This property is foundational for how NMF works and the dimensions are much smaller than the origina matrix $V$. Computing matrices $W$ and $H$ can be done in a variety of ways with a popular method being a multiplicative update method\cite{10.5555/3008751.3008829}.

\section{Contribution}

Using elements from non-negative matrix factorization, we have created a solution that attempts to find more optimal centroids for a cluster. Those centriods are then given to a classical clustering algrorithm such as $k$-means for partitioning into clusters. 

\subsection{Problem Formulation}

We formulate our QUBO from the input data which is heavily inspired by non-negative matrix factorization \cite{KimandPark-NMFClustering},\cite{bauckhage2015}. 
One drastic change is that our formulation will allow both positive and negative real numbers.\cite{Glover2018} In non-negative matrix factorization, we assume $V = WH$ where V the product of the two matrices $W$ and $H$ and both $V$ and $H$ contain only non-negative real numbers. Non-negative matrix factorization has an inherent cluster property where it will automatically cluster columns of input data $V = (v_1, ..., v_n)$. The approximation of $V$ via $V \cong WH$ is obtained by finding $W$ and $H$ that minimize the error function $||V-WH||_F$, subject to $W \geq 0, H \geq 0$ and $F$ being the Frobenius or $L^2$-norm. 
To encode our problem as a QUBO, we first created a series of \textit{substitution variables} that represent different combinations of unknown variables for our $W$ and $H$ matrices. \\

To better illustrate, let $V$ be a 2x2 matrix of real values with the shape of $p$x$n$. Let $W$ be a matrix also of 2x2 whose shape is $n$x$k$ and let $H$ be a 2x2 matrix with the shape of $k$x$n$. 
\[
    \begin{bmatrix}
        v_{11} & v_{12} \\
        v_{21} & v_{22} \\
    \end{bmatrix}\\ 
    \approx
        \begin{bmatrix}
        w_{11} & w_{12} \\
        w_{21} & w_{22} \\
    \end{bmatrix}\\ 
        \begin{bmatrix}
        h_{11} & h_{12} \\
        h_{21} & h_{22} \\
    \end{bmatrix}\\ 
\]\\
When we combine the right hand side, we end up with the following $WH$ matrix that is 2x2:

\[
WH = 
    \begin{bmatrix}
        w_{11}h_{11} + w_{12}h_{21} & w_{11}h_{12} + w_{12}h_{22} \\
        w_{21}h_{11} + w_{22}h_{21} & w_{21}h_{12} + w_{22}h_{22} \\
    \end{bmatrix}\\ \\
\]\\

When we apply the Frobenius norm $||V-WH||_2^2$, we have the following result:

\[
    \begin{bmatrix}
        (v_{11} - w_{11}h_{11} + w_{12}h_{21})^2 & (v_{12}-w_{11}h_{12} + w_{12}h_{22})^2 \\
        (v_{21} - w_{21}h_{11} + w_{22}h_{21})^2 & (v_{22} - w_{21}h_{12} + w_{22}h_{22})^2 \\
    \end{bmatrix}
\]\\

From this result, we can get the following quadratic equation

\begin{equation}
\resizebox{.999\hsize}{!}{
    $\bigg( \sqrt{
    (v_{11} - w_{11}h_{11} + w_{12}h_{21})^2 
    + (v_{12}-w_{11}h_{12} + w_{12}h_{22})^2 
    + (v_{21} - w_{21}h_{11} + w_{22}h_{21})^2 
    + (v_{22} - w_{21}h_{12} + w_{22}h_{22})^2}  \bigg)^2$}
\end{equation}

Upon expansion, we can see how the terms interact:

\begin{align*}
v_{11}^2+v_{12}^2+v_{21}^2+v_{22}^2-2 h_{11} v_{11} w_{11}-2 h_{12} v_{12} w_{11}+ h_{11}^2 w_{11}^2+h_{12}^2 w_{11}^2 \\
-2 h_{21} v_{11} w_{12}-2 h_{22} v_{12} w_{12}+2 h_{11} h_{21} w_{11} w_{12}+2 h_{12} h_{22} w_{11} w_{12}+h_{21}^2 w_{12}^2 \\
+h_{22}^2 w_{12}^2-2 h_{11} v_{21} w_{21}-2 h_{12} v_{22} w_{21}+h_{11}^2 w_{21}^2 +h_{12}^2 w_{21}^2-2 h_{21} v_{21} w_{22} \\
-2 h_{22} v_{22} w_{22}+2 h_{11} h_{21} w_{21} w_{22}+2 h_{12} h_{22} w_{21} w_{22}+h_{21}^2 w_{22}^2+h_{22}^2 w_{22}^2
\end{align*}

From this expansion we can derive our \textit{linear} coefficients and our \textit{quadratic} coefficients. Any constants that are created as part of the expansion are simply ignored in this formulation. The linear coefficients are those that correspond to a single variable, such as $v_{11}^2$ which would simply have the coefficient of $1$. Our quadratic coefficients are those that are paired along side multiple variables, such as $h_{12}v_{21}^2$. We do also end up with a coefficient paired along with $3$ or more variables, such as $-2h_{22}v_{22}w_{22}$. In the event we have $3$ or more variables, we must do a \textit{linearization} of the quadratic variables which is also known as a penalty function which will be described in the next section. One note that also needs to be mentioned, is since our variables are binary in nature which makes them \textit{idempotent}. Any squared variable is equal to the non-squared variable, for example: $v_{11}^2 = v_{11}$.

\subsection{Penalty Functions}

One of the challenges that we face is how to make sure that we pick the right center point consistently. Recall that we have our $H$ matrix above which is our "indicator" matrix that specifies which point will be the appropriate centroid. In our QUBO formulation, we can introduce what is known as a \textit{penalty function}. As the data is being computed in the annealing processes, the penalty function puts extra \textit{weight} on a solution that is less than optimal. With a default QUBO formulation, we rely on the classical and predictable expansion similar to $(a + b)^2 = a^2 + 2ab + b^2$.

 When we have our larger QUBO expression that is expanded in combination with our binary approximation, we can see that mathematically, we already have a penalty function \textit{implicitly} defined. This occurs when we mathematically have two quadratic variables that are multiplied with each other. For example, after the expansion of our quadratic equation we have $w1_{11}h_{11}$ somewhere in the equation, the penalty:
 \begin{equation}
     \delta_1(2(w_{11}h_{11} - 2(w_{11}h_{11})(w_{11} + h_{11}) + 3(w_{11}h_{11}))
 \end{equation}
 is applied\cite{cryptoeprint:2021:620} While this is beneficial for our work, it leaves open one problem. In our first example problem, we do not apply any additional penalties for our $H$ matrix. In some cases this may not be a problem, but we come across the potential that our $H$ matrix may have more than one centroid selected. To ensure we only have a single centroid selected per cluster, we introduce a second penalty function. Further referencing our quadratic expansion, we can apply the penalty:
 \begin{equation}
     \delta_2(1-\sum_{i=1}^k h_{ij})^2
 \end{equation}
One thing to note is that for each penalty function, we have $\delta_1$ and $\delta_2$ respectively. These are known formally as \textit{Lagrange multipliers}\cite{biz_calc}.

\subsection{Solution Formulation}

With our formulation defined, we submit our QUBO to a specified solver. The solver works using an adiabatic process over a period of time and during that time our unknown variables are replaced by computed values from the specific solver. One of the major benefits is that the solver is examining a much larger range of combinations and the most correct solution is the one that corresponds to the lowest energy value. When we get our result, we are returned a result that is series of binary values. These binary values read together are the result with the left most qubit being designated the \textit{sign qubit} and the remaining qubits used to represent the value \cite{borle2018analyzing}. In a similar fashion to Borle and Lomonaco, we also use a radix-2 approximation of the binary value which results in only supporting integers for the coordinates of a centroid.

To further formulate our problem, we can use the radix-2 approximation as outlined in a quantum linear least squares analysis by Borle and Lomonaco\cite{borle2018analyzing}. We can represent this a bit better visually by borrowing our $W$ matrix from the previous section:

\begin{center}
\begin{blockarray}{*{5}{c} l}
    \begin{block}{*{5}{>{$\footnotesize}c<{$}} l}
      8 & 4 & 2 & 1 \\
    \end{block}
    \begin{block}{[*{5}{c}]>{$\footnotesize}l<{$}}
      $W_{11}$ & $\hdots$ & $\hdots$ & $W_{1k}$ \\
      $\hdots$ & $\hdots$ & $\hdots$ & $\hdots$ \\
      $W_{n1}$ & $\hdots$ & $\hdots$ & $W_{nk}$ \\
    \end{block}
  \end{blockarray}
\end{center}

It will be assumed that the "endianess" for this part of the formulation will be from right to left, with the most significant "bit" on the left. Similar to Borle and Lomonaco \cite{borle2018analyzing}, we will also be using the next power of 2 combined with the next qubit as a "sign bit" so that we can represent positive and negative real numbers. The importance of this is that these will act as multipliers for the coefficients that are determined through quantum annealing. 

For a matrix that is using 4 total qubits including the most significant as our sign bit, our expanded form would look similar to $-8q_8 + 4q_4 + 2q_2 + q_1$ where $n$ in $q_n$ refers to having $n$ power of 2 as a label to our qubits. We can then express this more concisely: 

\begin{equation}
    \vartheta q_{j\theta} + \sum_{\theta \in \Theta}\theta q_{j\theta} = 
    \sum_{\theta \in \Theta} 2^{\theta}q_{j\theta} 
\end{equation}
where 
\[
\vartheta = 
\begin{cases}
-2^{p+1}, & two's\ complement \\
-2^{p+1} + 2^{\Theta} & one's\ complement \\
\end{cases}
\]\\
and $\Theta$ is the maximum power of 2 based on the number of qubits necessary. As an example, if $[o,p] = [0,2]$ and we need to represent $3$, then $q_{j4} = 0, q_{j2} = 1, q_{j1} = 1$ which gives us $w = 2^{2}q_{j4} + 2^{1}q_{j2} + 2^{0}q_{j1} = 3$.

If we continue using matrix $W$ as our example and combine our radix-2 approximation, we can expand each $w \in W$ to its binary equivalent:

\begin{equation}
 \begin{bmatrix}
    \big[-8w_{11}^{0}\ 4w_{11}^{4}\ 2w_{11}^{2}\ w_{11}^{1}\big] & \cdots & \big[-8w_{1k}^{0}\ 4w_{1k}^{4}\ 2w_{1k}^{2}\ w_{1k}^{1}\big] \\
    \cdots   & \cdots & \cdots \\
    \big[-8w_{n1}^{0}\ 4w_{n1}^{4}\ 2w_{n1}^{2}\ w_{n1}^{1}\big] & \cdots & \big[-8w_{nk}^{0}\ 4w_{nk}^{4}\ 2w_{nk}^{2}\ w_{nk}^{1}\big]
 \end{bmatrix} 
\end{equation}\\
where $w \in \big[-8w_{nk}^{0}\ 4w_{nk}^{4}\ 2w_{nk}^{2}\ w_{nk}^{1}\big]$ and $w^{n}$ refers to the specific qubit being utilized. In the case of NMF in the paper by Park\cite{KimandPark-NMFClustering}, we would typically do this same process for matrices $W$ and $H$ which represent our unknown values. When we have our resulting matrices $W$ and $H$ with our solved unknown variables, we can multiply $W$ and $H$ back together forming $WH$. Each column of $WH$ will then contain our centroids to be used for prototype based clustering.

\section{Results}

Using the free developer account provided by D-Wave, we have a monthly time limit we are allowed as such many of our sample sizes are limited in scope. Our results are utilizing random Gaussian blobs utilizing \textit{make\_blobs} from scikit-learn. We also ran our experiments using the MOTIF data set, which is a fairly large data set consisting of malware metadata primarily used for machine learning. \cite{Joyce2023} We also use and compare between three different processes provided by the D-Wave Ocean SDK: TABU, Simulated Annealing, and D-Wave's Hybrid BQM solver. \footnote{\url{https://www.dwavesys.com/resources/white-paper/hybrid-solvers-for-quadratic-optimization}} MOTIF is a multi-dimensional data set so to reduce our data to two usable dimensions, we used principal component analysis (PCA). To measure our clustering performance, we used a variety of common metrics utilized in cluster analysis: inertia, silhouette score, homogeneity, completeness and v-measure. As an additional metric, we also recorded the number of overall iterations $k$-means took when using the centroids. For cluster inertia and iterations, the lower value is more optimal where for silhouette, homogeneity, completeness, and v-measure a higher value is more optimal. We had an upper bound of the number of iterations allowed by $k$-means set to $10000$. \\

\subsection{Cluster Analysis Metrics}
The following is a list of these metrics and their definitions as it pertains to cluster analysis.

\begin{description}
\item[\textbf{Inertia: }] Sum of squared errors that pertains to the cluster. One generally wants this value to be as low possible. \\

\item[\textbf{Silhouette Score: }]  Also known as the \textit{Silhouette Coefficient}, a metric used to calculate the qualify of a clustering algorithm. Uses values ranged from $-1$ to $+1$.\\

\item[\textbf{Homogeneity: }] Used to measure how similar each of the samples are in a cluster. Uses values $-1$ to $1$.\\

\item[\textbf{Completeness: }]  Measures how similar samples are placed together by the clustering algorithm. Uses values $-1$ to $1$.\\

\item[\textbf{V-Measure: }] Also known as the \textit{Normalised Mutual Information Score}, is the average of homogeneity and completeness scores.\\
\end{description}

\subsection{Computational Processes}

The following is a list of the quantum and quantum inspired computational processes that were used. \\

\textbf{TABU Solver: } TABU search is a classical metaheuristic search method used in mathematical optimization. Local searches use a potential solution and checks the solutions immediate neighbors for a better solution. The TABU solver provided by D-Wave uses the MST2 multistart TABU search algorithm for problems expressed as quadratic unconstrained binary optimization formulations.\footnote{\url{https://docs.ocean.dwavesys.com/_/downloads/tabu/en/latest/pdf/}}.\\

\textbf{Simulated Annealing: } Simulated annealing is a classical probabilistic method for approximating a global optimum of a given function. Methods like simulated annealing are used when it's more important to approximate a global optimum than finding an exact local optimum. The simulated annealer that comes with the D-Wave software development kit is a classical approximation technique for approximating a global optimum. \\

\textbf{Hybrid BQM: } The D-Wave Hybrid BQM is a solver that combines the computing power of a quantum annealer (QPU) with a classical CPU. To use the Hybrid BQM, one signs up with their cloud platform known as LEAP\footnote{\url{https://cloud.dwavesys.com/leap/}}. The solver itself decides what parts of the problem are better suited by the QPU vs the CPU and schedules the problems accordingly. Due to the availability constraints in addition to our formulation constraints, we were only able to reliably use the Hybrid BQM for this formulation for up to 35 samples. We did manage to get a very close result for 40 samples, which we include for the sake of completeness.

\subsection{Gaussian Centroids}

Using random Gaussian centroids, we chose a seed value of $0$ and a cluster size $k$ of $3$. The figures below show a comparison between random centroids and those generated from QOCI utilizing the TABU, simulated annealing, and Hybrid BQM processes. 

\setlength\belowcaptionskip{-3ex}
\begin{figure}[H]
\centering
\includegraphics[scale=0.5]{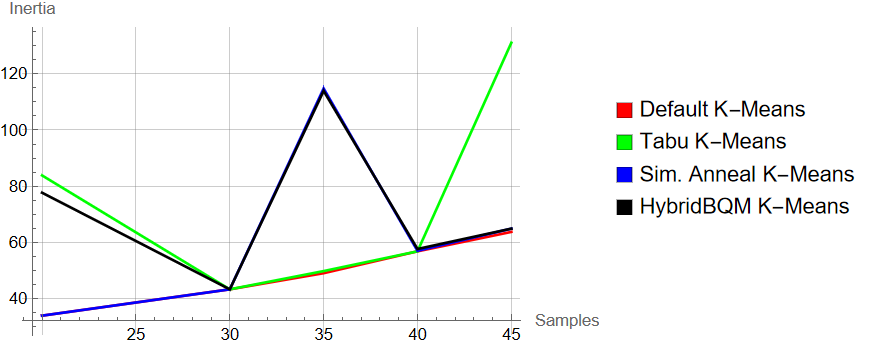}
\caption{Gaussians: Inertia}
\end{figure}

\begin{figure}[H]
\centering
\includegraphics[scale=0.5]{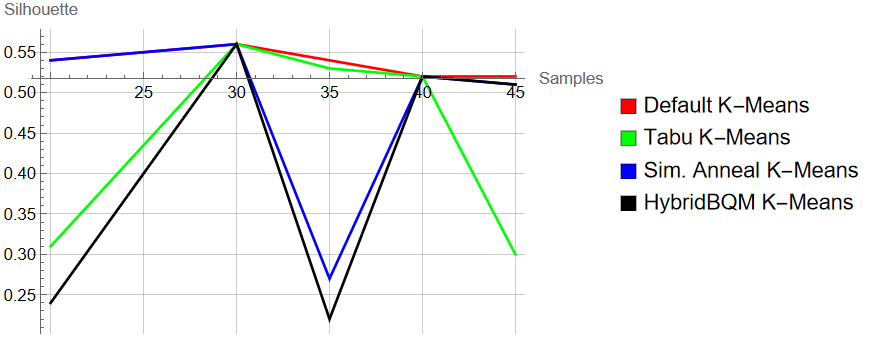}
\caption{Gaussians: Silhouette Scores}
\end{figure}

\begin{figure}[H]
\centering
\includegraphics[scale=0.5]{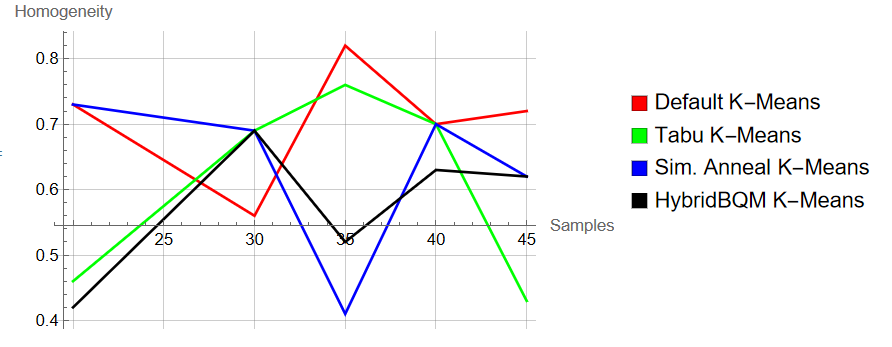}
\caption{Gaussians: Homogeneity Scores}
\end{figure}

\begin{figure}[H]
\centering
\includegraphics[scale=0.5]{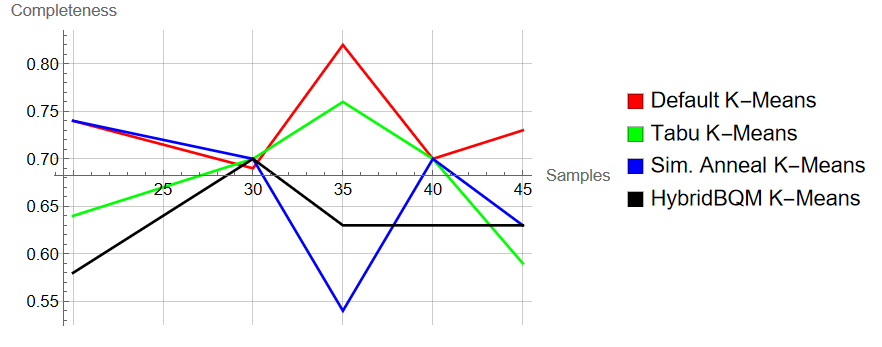}
\caption{Gaussians: Completeness Scores}
\end{figure}

\begin{figure}[H]
\centering
\includegraphics[scale=0.5]{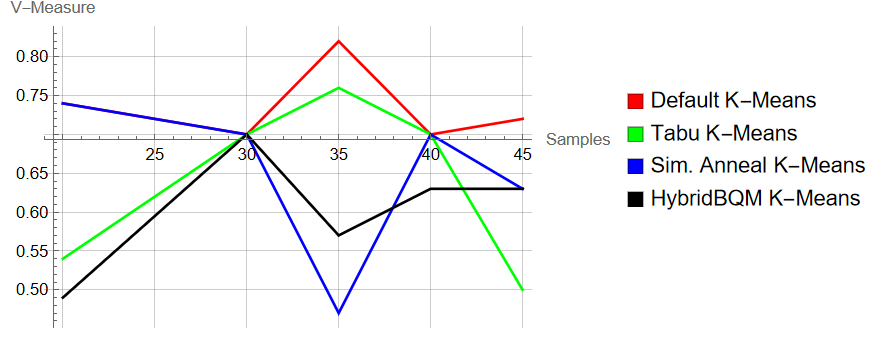}
\caption{Gaussians: V-Measure Scores}
\end{figure}

\begin{figure}[H]
\centering
\includegraphics[scale=0.5]{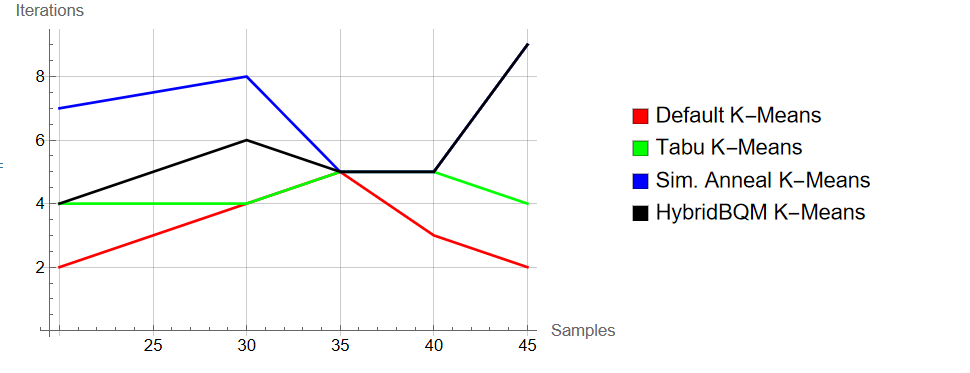}
\caption{Gaussians: Iterations}
\end{figure}

\subsection{MOTIF Centroids}

Using randomly chosen data from the MOTIF data set, we chose a seed value of $0$ and a cluster size $k$ of $3$. The figures below show a comparison between random centroids and those generated from QOCI utilizing the TABU, simulated annealing, and Hybrid BQM processes.

\begin{figure}[H]
\centering
\includegraphics[scale=0.5]{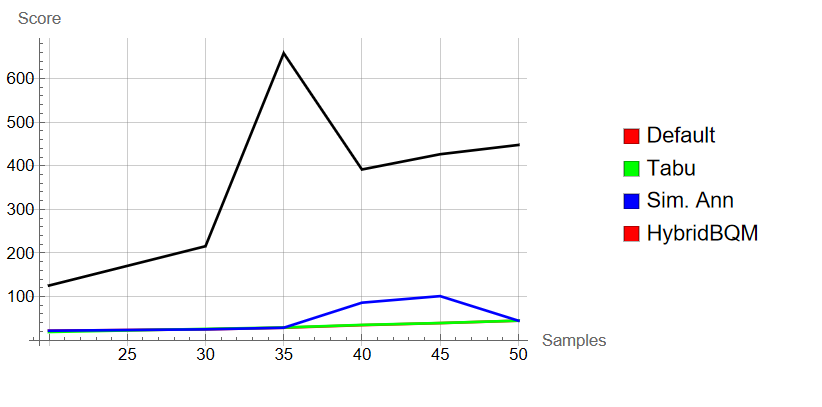}
\caption{MOTIF: Inertia}
\end{figure}

\begin{figure}[H]
\centering
\includegraphics[scale=0.5]{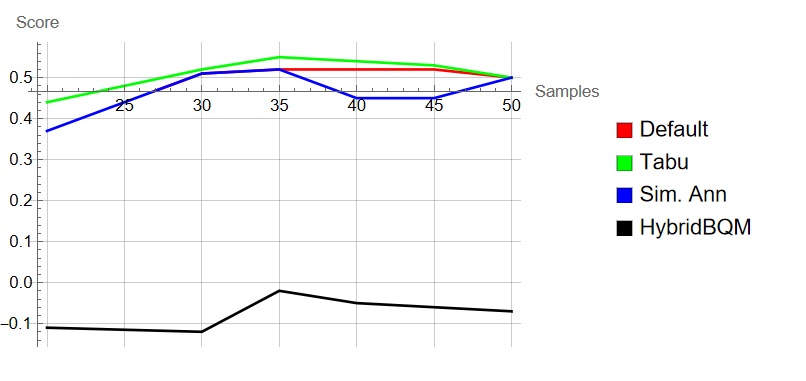}
\caption{MOTIF: Silhouette Score}
\end{figure}

\begin{figure}[H]
\centering
\includegraphics[scale=0.5]{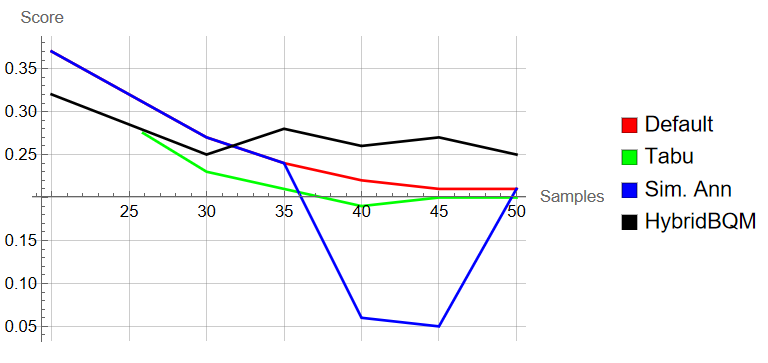}
\caption{MOTIF: Homogeneity Score}
\end{figure}

\begin{figure}[H]
\centering
\includegraphics[scale=0.5]{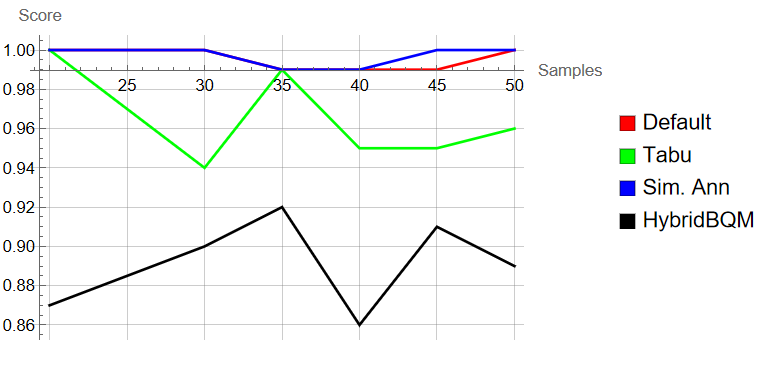}
\caption{MOTIF: Completeness Score}
\end{figure}

\begin{figure}[H]
\centering
\includegraphics[scale=0.5]{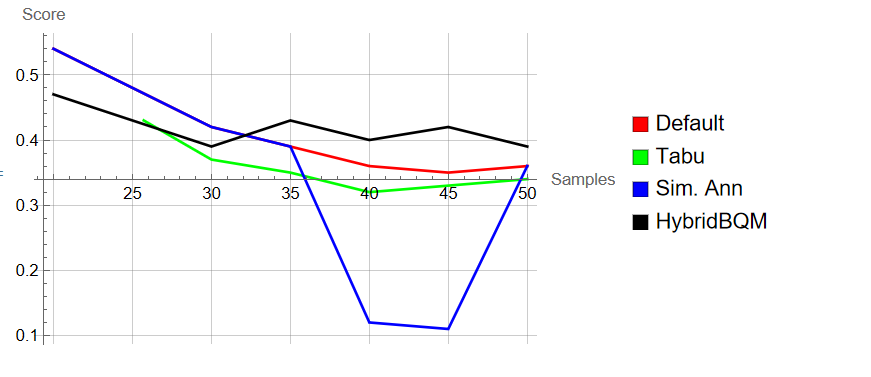}
\caption{MOTIF: V-Measure}
\end{figure}

\begin{figure}[H]
\centering
\includegraphics[scale=0.5]{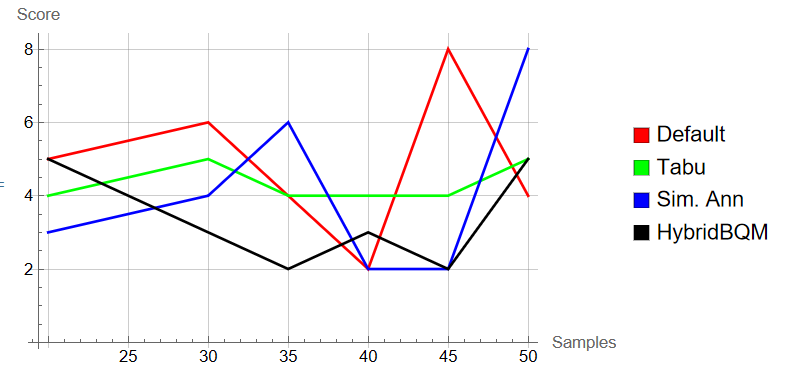}
\caption{MOTIF: Iterations}
\end{figure}

\section{Conclusion}

To summarize our findings for the Gaussian data, all processes for cluster inertia were nearly the same with the exception of the TABU solver which proved to be the least optimal. For the silhouette score, the random centroids along with Hybrid BQM were both similar and the more optimal processes. For the remaining scores of homogeneity, completeness, v-measure and iterations, the random centroids were more optimal.\\

Summarizing our findings for the MOTIF data set, the inertia was similar on all processes except the Hybrid BQM which was the least optimal. This was also true when comparing the silhouette score. For the homogeneity score the Hybrid BQM process was more optimal. The completeness score was quite close only between simulated annealing and Hybrid BQM. With v-measure, the Hybrid BQM again was the more optimal process. In terms of number of iterations taken by $k$-means, random, TABU, and Hybrid BQM all were similar in terms of number of iterations while simulated annealing had the highest number of iterations. \\

Quantum computing has a long way to go, however, it has come a long way in a very short period of time, especially when compared to computation as a whole. There are many avenues and popular places where quantum computing is being used, but it's important to take a step back and re-evaluate some of our existing work. For example, many combinatorial problems are NP-hard and we have classical heuristics to get us to a point where things are \textit{good enough}. These heuristics often are very good approximations, but as things evolve, good enough may no longer be good enough.

%
\bibliographystyle{unsrt} 
\bibliography{references.bib}

\begin{thebibliography}{10}

\bibitem{algorithms_clustering_data}
Dubes R.~C. Jain A.~K.
\newblock {\em Algorithms for Clustering Data.}
\newblock Prentice-Hall, 1988.

\bibitem{MacQueen_clustering}
J.~B. MacQueen.
\newblock Some methods for classification and analysis of multivariate
  observations.
\newblock In {\em Proceedings of 5th Berkeley Symposium on Mathematical
  Statistics and Probability: 281-297}. University of California Press, 1967.

\bibitem{weber_prob}
Hansen et~ala. Chen, Pey-Chun.
\newblock Weber's problem with attraction and repulsion.
\newblock {\em Journal of Regional Science}, 32:467--486, 1992.

\bibitem{dwave}
D-Wave.
\newblock {D-wave}.
\newblock \url{https://dwavesys.com}, 2020.

\bibitem{10.5555/3008751.3008829}
Daniel~D. Lee and H.~Sebastian Seung.
\newblock Algorithms for non-negative matrix factorization.
\newblock In {\em Proceedings of the 13th International Conference on Neural
  Information Processing Systems}, NIPS'00, page 535–541, Cambridge, MA, USA,
  2000. MIT Press.

\bibitem{KimandPark-NMFClustering}
Jingu Kim and Haesun Park.
\newblock Sparse nonnegative matrix factorization for clustering.
\newblock Technical Report GT-CSE-08-01, Georgia Institute of Technology, 2008.

\bibitem{bauckhage2015}
Christian Bauckhage.
\newblock k-means clustering is matrix factorization, 2015.
\newblock \url{https://arxiv.org/abs/1512.07548}.

\bibitem{Glover2018}
Fred~W. Glover and Gary~A. Kochenberger.
\newblock A tutorial on formulating {QUBO} models.
\newblock {\em CoRR}, abs/1811.11538, 2018.

\bibitem{cryptoeprint:2021:620}
Elżbieta Burek, Michał Misztal, and Michał Wroński.
\newblock Algebraic attacks on block ciphers using quantum annealing.
\newblock Cryptology ePrint Archive, Report 2021/620, 2021.
\newblock \url{https://ia.cr/2021/620}.

\bibitem{biz_calc}
Laurence~D. Hoffmann and Gerald~L. Bradley.
\newblock {\em Calculus for Business, Economics, and the Social and Life
  Sciences (8th ed.)}.
\newblock McGraw Hill, 2004.

\bibitem{borle2018analyzing}
Ajinkya Borle and Samuel~J. Lomonaco.
\newblock Analyzing the quantum annealing approach for solving linear least
  squares problems.
\newblock {\em arXiv 1809.07649}, 2018.

\bibitem{Joyce2023}
Robert~J. Joyce, Dev Amlani, Charles Nicholas, and Edward Raff.
\newblock {MOTIF}: A large malware reference dataset with ground truth family
  labels.
\newblock {\em Computers and Security}, 124, Issue C, January 2023.

\end{thebibliography}

\end{document}